# Prospects for long-range reactor monitoring with gadolinium-loaded water-Cherenkov detectors

MICHAEL LEYTON[1,*] and STEPHEN DYE[2]

[1] Institut de Física d'Altes Energies (IFAE), Barcelona Institute of Science and Technology
[2] Department of Physics and Astronomy, University of Hawaii
*Corresponding author. E-mail: leyton@cern.ch

**Abstract.** Antineutrino detectors are practical, non-intrusive tools capable of remotely monitoring the activity of nuclear reactors. Here we explore the sensitivity of the Super-Kamiokande water-Cherenkov detector, following gadolinium loading, to antineutrinos from a nuclear reactor complex at a distance of approximately 190 km. The livetimes required to observe the two currently operating cores in the reactor complex depend on the activity of other reactors in the vicinity, as well as on estimates of detection efficiency and background rates. Under reasonable assumptions, we find that gadolinium-loaded Super-Kamiokande could detect the flux of antineutrinos from both cores at the Takahama reactor complex at 95% confidence level in 50 (10) live days 95% (50%) of the time, or the flux from one core in 397 (73) live days, provided that each core is operating at nominal power.

## 1. Introduction

Reactor safeguards aim to detect the diversion of fissile materials from civil nuclear reactor facilities into weapons programs. In comparison to current safeguards, which rely on bookkeeping and surveillance, reactor monitoring with antineutrino detectors can provide a more direct, and less intrusive, way to measure the operation of reactors and the evolution of their fuel. A strategic goal of reactor monitoring with antineutrinos is, therefore, to remotely detect a change in the operational status of a reactor.

Water-Cherenkov detectors are well suited for this purpose since the technology is scalable to the very large target masses (~Mtonne) needed for long-range (>100 km) reactor monitoring. The Super-Kamiokande neutrino observatory [1] is a water-Cherenkov detector located 1 km underground at the Kamioka Observatory in Japan, with a fiducial mass of 22.5 ktonne. It will soon undergo a significant upgrade, adding a small fraction (~0.1% by mass) of gadolinium (Gd) to its ultra-pure water target [2]. The addition of Gd to the water target allows neutrons to be easily detected and low-energy antineutrinos to be efficiently identified, providing a significant boost in sensitivity to neutrino-induced inverse beta decay reactions ($\bar{\nu}_e p \to e^+ n$).

Due to the Fukushima-Daiichi disaster in 2011, only a handful of reactors in Japan are currently in operation: two at the Sendai reactor complex (829 km from Kamioka); one at Ikata (559 km from Kamioka); and two at Takahama (191 km from Kamioka). This minimal level of reactor activity[1] in the vicinity of Kamioka provides a unique window in which to demonstrate the capability of the upgraded Super-Kamiokande detector, here referred to as SuperK-Gd, to monitor reactors at long range.

In this paper, we examine the prospects of SuperK-Gd to monitor cores at Takahama, the reactor complex closest to Kamioka currently in operation. We calculate the livetime needed for SuperK-Gd to detect the flux of antineutrinoss produced by reactors at Takahama at 95% confidence level. The Takahama-Kamioka baseline distance of 191 km is approximately an order of magnitude larger than that of proposed reactor monitoring demonstration experiments using Gd-doped water-Cherenkov detectors [3], strengthening the argument, and widening the possible use cases, for reactor monitoring with antineutrinos.

---

[1]Since preparation of the analysis, it was announced that two reactors at Ohi will restart in 2018. Reactors at Ohi are not considered here, but will be studied in a future publication.

## 2. Antineutrino spectra

Nuclear power reactors generate heat via neutron-induced fissions of uranium (U) and plutonium (Pu) and the subsequent decays of unstable byproducts. The fission fragments are produced in highly excited states that decay principally by beta ($\beta^-$) decay, emitting ~6 antineutrinos ($\overline{\nu}_e$) per fission on average. In a typical reactor, more than 99.9% [4] of $\overline{\nu}_e$s above the threshold of inverse beta decay on free protons (1.806 MeV) originate in the fission process of four isotopes: $^{235}$U, $^{238}$U, $^{239}$Pu and $^{241}$Pu.

We construct $\overline{\nu}_e$ emission spectra for each of the four principal contributing isotopes using datasets and fits from ILL [5, 6, 7, 8], Mueller et al. [9], Huber [10] and Vogel et al. [11, 12]. Details of the calculation will be published at a later date. The final $\overline{\nu}_e$ spectra for each of the isotopes, with error bands, are shown in Fig. 1.

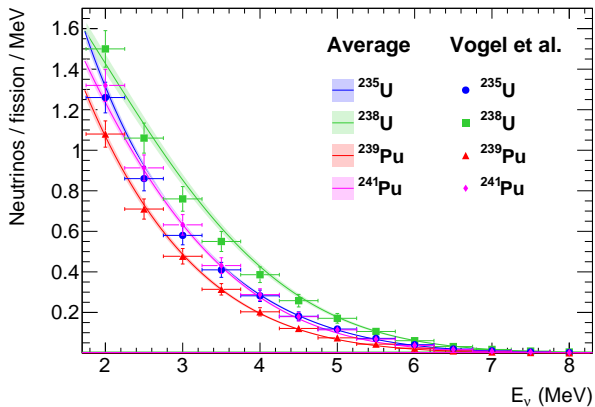

**Figure 1.** **Reference isotopic $\overline{\nu}_e$ spectra per fission**, shown with $1\sigma$ error band, for $^{235}$U (blue), $^{238}$U (green), $^{239}$Pu (red) and $^{241}$Pu (magenta). Also shown for comparison are $\overline{\nu}_e$ spectra from Vogel et al. [11, 12] for $^{235}$U (blue circles), $^{238}$U (green squares), $^{239}$Pu (red triangles) and $^{241}$Pu (magenta diamonds).

We then construct reactor $\overline{\nu}_e$ emission spectra for three categories of reactors, grouped according to their cooling and moderating materials: I) pressurized water reactors (PWRs), boiling water reactors (BWRs), fast breeder reactors (FBRs), light water-cooled graphite-moderated reactors (LWGR) and gas-cooled reactors (GCRs); II) pressurized heavy-water reactor (PHWRs); or III) reactors burning mixed oxide fuel (MOX), assumed to provide 30% of the total power, while the remaining 70% is produced by standard (category I) fuel. The reactor $\overline{\nu}_e$ emission spectra are proportional to the total thermal power output ($P_{\text{th}}$) of the reactor and the fraction of the total thermal power produced by each of the four contributing isotopes, summed over all four isotopes. Although power fractions are generally time-dependent quantities that evolve over the reactor operation cycle, here we assume power fractions [13, 14] that correspond to the midpoint of the operation cycle for simplicity. Fig. 2 shows the calculated $\overline{\nu}_e$ emission spectra for the three reactor categories considered here, per unit of thermal power output.

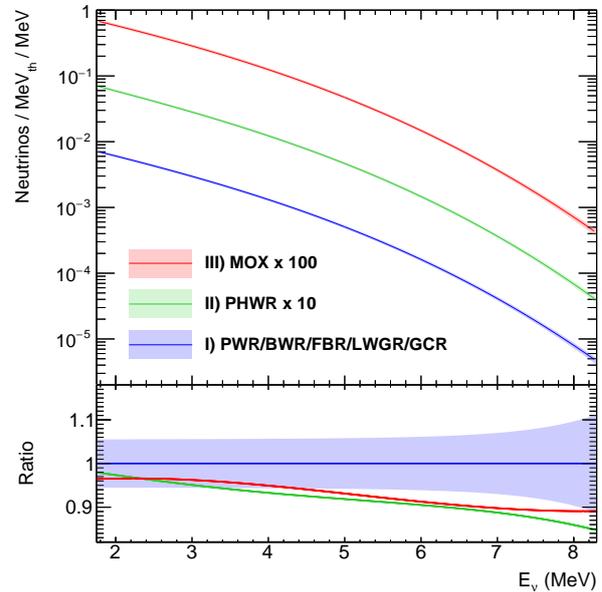

**Figure 2.** **Reactor $\overline{\nu}_e$ emission spectra per unit of thermal power output** for I) pressurized water (PW), boiling water (BW), fast breeder (FB), light water-cooled graphite-moderated (LWG) and gas-cooled (GC) reactors (red); II) pressurized heavy-water (PHW) reactors (blue); and III) reactors burning mixed oxide fuel (MOX). Error bands include uncertainties on isotope spectra, thermal energy released per fission and a ±5% uncertainty on mid-cycle power fractions from [13, 14]. The bottom plot shows the $1\sigma$ error band (blue) on the $\overline{\nu}_e$ emission spectrum for category I reactors and the ratio of $\overline{\nu}_e$ emission spectra for category I (blue), II (green) or III (red) reactors over that for category I reactors.

## 3. SuperK-Gd simulation

We simulate the expected performance of the upgraded SuperK-Gd detector, based on existing data from the fourth phase of Super-Kamiokande (SK-IV) and taking into account the impact of adding 0.2% (by mass) gadolinium sulfate, $Gd_2(SO_4)_3$, to the ultra-pure water.

Gd has a neutron capture cross section ~5 orders of magnitude larger than protons and emits a gamma ($\gamma$) cascade of 8 MeV, allowing neutrons to be easily detected and low-energy $\overline{\nu}_e$s to be efficiently identified via the delayed coincidence of neutrons produced in inverse beta decay reactions. The energy spectrum of the $\gamma$ rays emitted by neutron capture on Gd has been measured in a test vessel to have a mean energy of $4.3 \pm 0.1$ MeV [15], in good agreement with the MC prediction.

The energy spectrum due to neutron capture on Gd is parameterized using a Gaussian distribution with the measured mean and width from [15]. For each event, a value at random is picked from the distribution to represent the total detected energy from $\gamma$s. The trigger efficiency from the Super Low Energy (SLE) trigger of SK-IV is then applied by fitting the values given in [16] (84% between 3.49 and 3.99 MeV, 99–100% above 3.99 MeV). The reconstructed positron kinetic energy $T_e^{\text{reco}}$ is calculated by smearing the total positron energy $E_e$ according to the resolution parameterization from SK-IV [16]: $\sigma(E_e) = -0.0839 + 0.349\sqrt{E_e} + 0.0397 E_e$.

All events with $T_e^{\text{reco}} < 2.5$ MeV are rejected and an error function is used to simulate the positron reconstruction efficiency, ranging from 80% at 3.0 MeV, up to 95% above 5.0 MeV. A Gd capture efficiency of 90% and neutron reconstruction efficiency of 90% are also applied [17]. Finally, an acceptance of 80% [17] is applied in order to simulate the effect of a spallation cut, used to remove $\beta$ and/or $\gamma$ decays of radioactive elements produced by cosmic-ray muons that break up an oxygen nucleus in the target.

We assume a fiducial volume of 22.5 ktonne [16] to reject events near the wall of the detector and ignore the small reduction in fiducial volume for low-energy events.

## 4. Interaction and event rates

We calculate the cross section of quasi-elastic neutrino-proton scattering, $\bar{\nu}_e p \rightarrow e^+ n$, also referred to as inverse beta decay (IBD), from the V-A theory of weak interactions, neglecting energy-dependent recoil, weak magnetism and radiative corrections [18]. Assuming the nucleon mass is infinite, then the energy of the incident neutrino $E_\nu$ relates to the energy of the positron $E_e$ by $E_\nu = E_e + \Delta$, where $\Delta = m_n - m_p$ is the neutron-proton mass difference [18].

The number of IBD interactions induced by a reactor with thermal power $P_{\text{th}}$, as a function of standoff distance $L$ and antineutrino energy $E_\nu$, is calculated as:

$$N(L, E_\nu) = \frac{n_p \tau P_{\text{th}}}{4\pi L^2} \int \sigma^{\text{IBD}}(E_\nu) \frac{dR}{dE_\nu} P_{e\rightarrow e}(L, E_\nu) dE_\nu, \quad (1)$$

where $n_p$ is the number of free protons (e.g. hydrogen nuclei), $\tau$ is the exposure time, $dR/dE_\nu$ is one of the reactor $\bar{\nu}_e$ emission spectra from Fig. 2, $P_{e\rightarrow e}(L, E_\nu)$ is the neutrino survival probability after traveling a distance $L$ and $\sigma^{\text{IBD}}(E_\nu)$ is the quasi-elastic neutrino-proton scattering cross section.

Thermal power outputs ($P_{\text{th}}$) of reactor cores currently in operation worldwide are taken from 2016 data from [4, 19], updated with three additional reactors that have come online since the beginning of 2017. The distance of each reactor core to Kamioka is calculated assuming a spherical Earth.

Figure 3 shows the expected interaction rate from a single reactor core ($P_{\text{th}} = 2660$ MW$_{\text{th}}$) at Takahama, as a function of true positron kinetic energy $T_e$, compared with that from all background nuclear reactor cores. The SuperK-Gd simulation, detailed in the previous section, is then applied on the interaction rates in order to calculate the expected event rates, shown in Fig. 3 as a function of reconstructed positron kinetic energy $T_e^{\text{reco}}$. The calculations here estimate 7.36 events/month between $2 < T_e^{\text{reco}} < 8$ MeV from a single core at Takahama and 12.92 events/month from all background nuclear reactors. These estimates assume that each reactor core is operating continuously at nominal power.

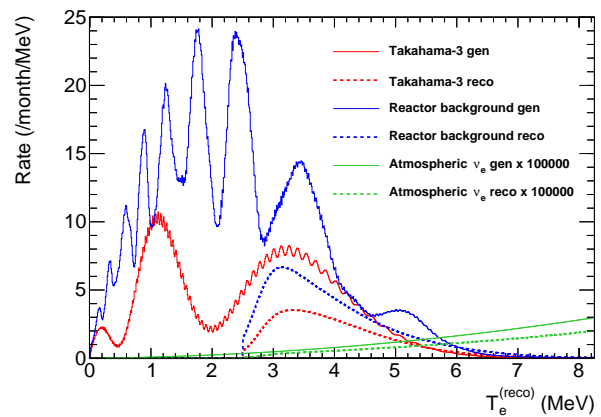

**Figure 3**. **Interaction and event rates** per month at SuperK-Gd for a single reactor core (2660 MW$_{\text{th}}$) at Takahama (red), compared with background from worldwide nuclear reactors (blue) and atmospheric $\bar{\nu}_e$s (green). Interaction rates (solid lines) from Equation 1 are calculated including the effect of neutrino oscillation and shown as a function of true positron kinetic energy $T_e$. Event rates (dashed lines) assume the detector performance discussed in Section 3 and are shown as a function of reconstructed positron kinetic energy $T_e^{\text{reco}}$. Atmospheric $\bar{\nu}_e$ rates have been scaled by a factor of 100,000.

## 5. Non-reactor backgrounds

In addition to the $\bar{\nu}_e$ background due to other reactors, both neutrino and non-neutrino sources can contribute to irreducible backgrounds in SuperK-Gd.

### 5.1 *Neutrino backgrounds*

We estimate the rates of IBD reactions due to geo-$\bar{\nu}_e$s, supernova relic neutrinos and atmospheric neutrinos via the charged current channel, as well as the rate of neutral current interactions from atmospheric neutrinos.

Geo-neutrinos are emitted by radioactive decays in the Earth's crust and mantle, but have energies below typical trigger thresholds of water-Cherenkov detectors. Using the predicted flux and spectra of geo-$\bar{\nu}_e$s at Kamioka from [20], we calculate the expected event rate at SuperK-Gd to be $< 2 \times 10^{-8}$/month.

Supernova relic neutrinos (SRNs), emitted by core collapse supernovae throughout the universe, are expected to contribute [17] 0.5–8.1 events/year between $9.5 < T_e < 29.5$ MeV at SuperK-Gd, depending on the theoretical model, and assuming 80% IBD detection efficiency. Scaling to the energy range of interest for this analysis, $2 < T_e < 8$ MeV, we conservatively estimate the background due to SRNs to be 0.15–2.43 events/year at SuperK-Gd, or $0.11 \pm 0.09$ events/month.

Charged current (CC) interactions of atmospheric $\bar{\nu}_e$s are indistinguishable from reactor $\bar{\nu}_e$ event-by-event, but their spectra are very different since the atmospheric neutrino flux is suppressed at low energies ($E_\nu \lesssim 500$ MeV) and rises with $E_\nu$. We estimate the background due to atmospheric $\bar{\nu}_e$ CC interactions at SuperK-Gd using 3D flux calculations from Bartol [21], averaged over all directions. Assuming average survival probability after oscillation, $P_{e \to e} = (1 - 0.5 \sin^2 2\theta_{12}) = 0.55$, we expect $< 0.0001$ events per month at SuperK-Gd for $T_e < 8$ MeV. Interaction and event rates for atmospheric $\bar{\nu}_e$s are shown in Fig. 3 as a function of $T_e^{\text{reco}}$.

Atmospheric $\nu_\mu$s can create a muon via a CC interaction, whose decay electron may not be correlated to the muon if the muon is below the Cherenkov threshold. However, the peak of the decay electron energy spectrum is closer to half the muon mass and the contribution below 8 MeV is expected to be negligible.

Neutrino-nucleus neutral-current (NC) interactions of atmospheric neutrinos pose an additional background to the IBD signal from reactor $\bar{\nu}_e$s. Such interactions can readily be identified in water-Cherenkov detectors since they produce nucleons or $\gamma$ rays originating from nuclear deexcitation. Quasi-elastic (QE) nucleon knockout, e.g. $\nu_x + {}^{16}O \to \nu_x + p + {}^{15}N^*$ or $\nu_x + {}^{16}O \to \nu_x + n + {}^{15}O^*$, becomes important for $E_\nu \gtrsim 200$ MeV [22]. The latter reaction, in particular, can emit a neutron with associated $\gamma$ rays with energies between 1 and 10 MeV [23], closely mimicking the IBD + Gd capture event signature.

To calculate the expected number of neutrino-nucleus interactions, we use the partial cross sections for neutral-current neutrino-induced reactions on ${}^{16}O$ calculated in [24] and the flux spectra of atmospheric neutrinos from Bartol [21]. The expected interaction rate at SuperK-Gd, summed over electron and muon flavors, is 0.055 events/month, or less than 0.035 events/month after applying Gd capture, neutron reconstruction and spallation cut efficiencies.

### 5.2 *Non-neutrino backgrounds*

To mimic the IBD + Gd neutron capture event signature, non-neutrino backgrounds must produce two flashes of Cherenkov light within $\sim 100\,\mu s$. Here we estimate the rate of non-neutrino backgrounds, including cosmogenic ${}^9$Li, fast neutrons, accidental coincidences and spontaneous fission of ${}^{238}$U.

Cosmic-ray-muon spallation byproducts can fake an inverse beta decay reaction if they decay and emit $\beta$s and $\gamma$s. The largest contributor to cosmic-ray-muon-induced background for our energy range of interest is ${}^9$Li, whose decay mode with $\beta^- n$ emission can produce detectable energy up to $\sim 12$ MeV. The background from cosmic-ray-muon-induced ${}^9$Li events was measured in SK-IV [17] down to $T_\beta = 7.5$ MeV. Following spallation cuts, the expected background in SuperK-Gd from cosmogenic ${}^9$Li, assuming 80% neutron tagging efficiency and 0.5% probability of muon-induced ${}^9$Li event leakage, is expected to be $0.5 \pm 0.1 \pm 0.2$ events/year between $9.5 < T_\beta < 29.5$ MeV [17]. We scale this prediction to our energy range of interest by calculating the theoretical $\beta^-$ spectrum of ${}^9$Li decay modes leading to $\beta^- n$ emission, applying positron threshold, trigger efficiency and energy resolution, as per the SuperK-Gd simulation described in Section 3, and scaling such that the integral of the region $9.5 < T_\beta < 29.5$ MeV is equal to 0.5 events/year. The estimated rate below 8 MeV at SuperK-Gd is $0.64 \pm 0.28$ events/month.

Cosmic ray muons can also produce fast neutrons when they traverse the detector or external rock of the underground cavern. Here we assume that neutrons produced by muons traversing the detector can be tagged effectively using the active veto and removed from the event sample. Although the outer detector volume can serve as a passive shield, neutrons produced in the external rock are more difficult to tag in coincidence with the primary muon due to the hard energy spectrum and long propagation range.

We estimate the rate of events due to cosmogenic fast neutrons that could fake the IBD event signature by producing two flashes of Cherenkov light within $\sim 100\,\mu s$, here referred to as 'di-neutron' events. We use predictions of the neutron flux and energy spectrum at the rock/cavern boundary at Kamioka by Mei & Hime [25]. Assuming a dome-shaped cavern with a diameter and height of 40 m, we estimate an incident rate of 16,590 neutrons/day, for $E_n > 1$ MeV. These neutrons are predominantly either caught by the 2.5-m-thick passive water shield (2 m of which is active veto) or reconstructed outside the fiducial volume. Scaling from calculations in [3], we estimate that only $37.5 \pm 8.6$ neutrons/month will create a prompt and delayed pair in the SuperK-Gd fiducial volume within $\sim 100\,\mu s$, of which $8.0 \pm 1.8$ neutrons/month will be reconstructed between 2 and 8

MeV and pass spallation cuts.

The rate of accidental coincidences is difficult to estimate properly without a full simulation since it depends on the composition of the detector and environment, vertex position in the detector and event selection criteria. Here we assume that the singles rate above threshold after all selection is ≲1 Hz, so that the rate of accidental coincidences can effectively be ignored, but a more detailed calculation is warranted.

Finally, we estimate the background due to spontaneous fission of U, resulting from the addition of $Gd_2(SO_4)_3$ to the ultra-pure water target [26]. Assuming a background of 0.11 events per year per mBq/kg of $Gd_2(SO_4)_3$ and an activity of 20 mBq/kg, we calculate an upper limit of 0.183 events/month at SuperK-Gd.

## 6. Sensitivity analysis

We calculate the livetime needed for SuperK-Gd to detect the flux of reactor $\bar{\nu}_e$s produced by one or two cores at Takahama at 95% confidence level (CL), for each of the following three scenarios:

A  two cores at Takahama operating at ≥ nominal power;

B  one core at Takahama operating at ≥ nominal power and the other core completely off (e.g. for maintenance or refueling);

C  same as scenario A, but one core treated as 'signal' and the other as 'background'.

All other (452) reactor cores currently in operation [4, 19] are treated as background and conservatively assumed to be operating at nominal power. Compiling monthly data from 2003–2010 from [19], we find that the two Takahama cores were operating in scenario A or C (B) for 47% (29%) of calendar months. For 21% of calendar months, one core was operating at ≥ nominal power and the other was partially on (nominal power < 100%), most likely due to refueling for a portion of the calendar month. For the remainder of calendar months (3%), one core was partially on and the other core was either partially on or completely off. For no calendar months were both cores entirely off.

We use the methodology developed in [20]. For each scenario and for a given livetime, we run a set of 1000 pseudo-experiments according to the event rates and spectra calculated in Sections 4 and 5. The number of events in each pseudo-experiment is fluctuated according to the statistical error on the predicted event rates and scaled to the total livetime of the experiment. For each set of pseudo-experiments, we use a profile likelihood statistic to calculate the 95% confidence intervals, assuming that the data contain both signal and background events. We then assess the livetime at which 95% (50%) of pseudo-experiments exclude the null (background-only) hypothesis at 95% CL. An example pseudo-experiment for scenario A is shown in Fig. 4.

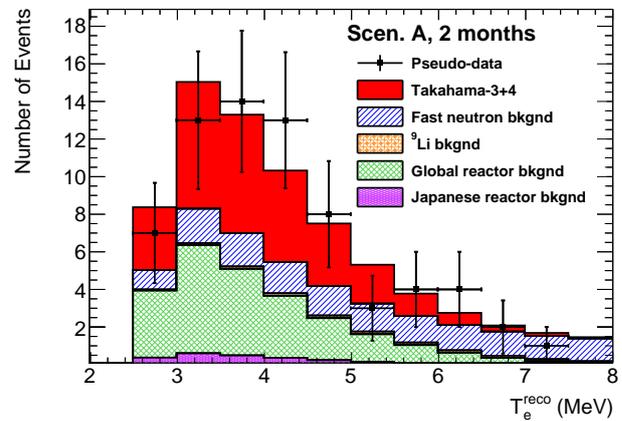

**Figure 4.** Positron kinetic energy spectrum from IBD interactions and correlated backgrounds at SuperK-Gd (scenario A). Signal events induced by $\bar{\nu}_e$s from both Takahama reactor cores (red solid) are shown together with background events induced by $\bar{\nu}_e$s from Japanese reactors (purple fine hatched) and all other global reactors (green hatched). All reactor cores are assumed to be operating continuously at nominal power. Di-neutron background due to cosmic-ray-induced fast neutrons (blue striped) and leakage of background from $^9$Li decaying via $\beta^-n$ (orange weave) are also shown. Pseudo-experimental data (black squares) are shown with statistical and systematic uncertainties added in quadrature. Distributions have been normalized to a livetime of 2 months, including the effect of neutrino oscillation.

Systematic uncertainties on both signal and background distributions are applied when calculating the likelihood statistic. To cover measurement uncertainties, we include the (symmetrized) energy-uncorrelated systematic uncertainties measured by SK-IV [16], which include uncertainties on trigger efficiency, reconstruction quality, event selection and cross section, decreasing from 5.0% below 4 MeV to 0.9% for 7.5–8.0 MeV. We also include uncertainties on energy scale (±0.54%) and energy resolution (±1.0%) from SK-IV [16].

Uncertainties on the predicted reactor spectra are mostly flat in energy, increasing gradually from 5.5%–6.0% for $1.75 < E_\nu < 5.5$ MeV, up to 7.1% for $7 < E_\nu < 8$ MeV, with a small dependence on reactor category. We apply an additional 2.5% uncertainty [4] on reactor-$\bar{\nu}_e$ fluxes uncertainties on oscillation parameters, conversion of reactor power to flux, power value reported by plant operators and seasonal changes in the reactor power output due to load requirements on the user side, refueling or long-term shutdown.

The results of the sensitivity analysis are shown in Fig. 5.

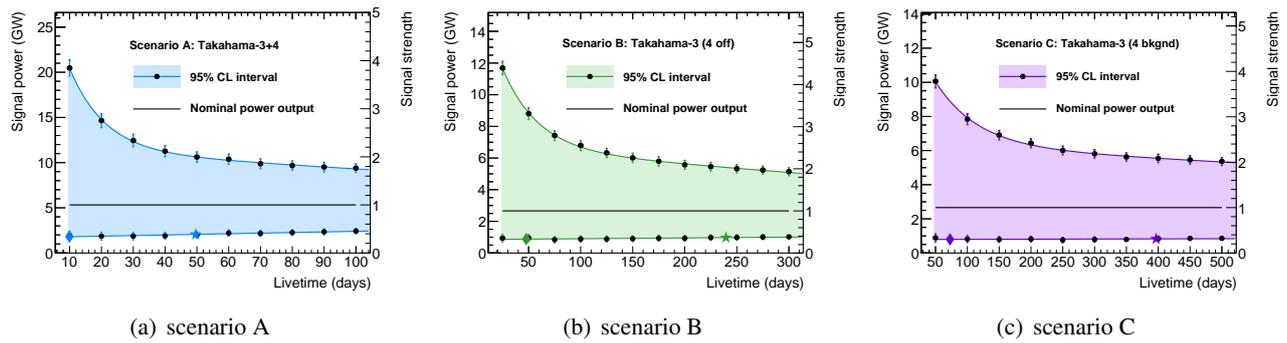

(a) scenario A  (b) scenario B  (c) scenario C

**Figure 5.** **Confidence interval at SuperK-Gd vs. livetime.** Mean 95% confidence interval for signal + background pseudo-experiments for three scenarios. In scenario A (a), both reactor cores at Takahama are on, operating at nominal power and treated as signal. In scenario B (b), one core is operating at nominal power and the other core is completely off. In scenario C (c), both cores are on and operating at nominal power, but one is treated as signal and the other as background. The star- (diamond-) shaped markers represent the exposures at which 95% (50%) of pseudo-experiments exclude the null (background-only) hypothesis at 95% confidence level. Each curve has been fit with a line plus exponential.

## 7. Conclusion

We study the sensitivity of Super-Kamiokande, following gadolinium loading, to antineutrinos from the Takahama reactor complex at a distance of approximately 190 km. Assuming modest detector performance and reasonable estimates of background rates, we find that gadolinium-loaded Super-Kamiokande could detect the flux of $\bar{\nu}_e$s from both cores operating at nominal power (2.66 GW$_{th}$ each) at 95% confidence level in 50 (10) live days 95% (50%) of the time. If one core is off for refueling or maintenance, the flux of antineutrinos from the other core operating at nominal power could be detected in 240 (48) live days at 95% confidence level 95% (50%) of the time. Finally, the flux of antineutrinos from a single core could be detected in the presence of antineutrinos from the other core at 95% confidence level in 397 (73) live days 95% (50%) of the time, assuming both cores are operating at nominal power. These results suggest that gadolinium-loaded water-Cherenkov detectors are capable of observing the operational status of a single core within a reactor complex of known power at a range of ∼200 km.

## Acknowledgements

M.L. acknowledges support from the Marie Skłodowska-Curie Fellowship program, under grant 665919 (EU, H2020-MSCA-COFUND-2014), Ministerio de Economia, Industria y Competitividad (MINECO), Agencia Estatal de Investigación (AEI) and Fondo Europeo de Desarrollo Regional (FEDER), under grants FPA2014-77347-C2-2 and SEV-2012-0234.